\documentclass[pre,twocolumn,groupedaddress]{revtex4-1}
\usepackage[utf8]{inputenc}
\usepackage{amsmath}
\usepackage{amsfonts}
\usepackage{amssymb}
\usepackage{graphicx}
\usepackage{graphics}
\usepackage{color}
\begin{document}
\title{Optimized Coefficient of performance of Power law dissipative Carnot like Refrigerator}  
\author{K. Nilavarasi$^1$  and  M. Ponmurugan$^2$}
\email[]{ponphy@cutn.ac.in}
\affiliation{$^1$ Department of Physics, National Institute of Technology-Thiruchirapalli,
Thiruchirapalli- 620 015, Tamil Nadu, India. email:nilavarasikv@gmail.com  \\
$^2$ Department of Physics, School of Basic and Applied Sciences, Central
University of Tamil Nadu, Thiruvarur 610 005, Tamil Nadu, India.}


\begin{abstract}
The present work investigates the  generalized  extreme bounds of the coefficient of performance (COP) for the power law dissipative Carnot-like refrigerator in the presence of non-adiabatic dissipation under $\chi$ and $\dot{\Omega}$ optimization criteria. The lower and upper bounds of the COP for the low dissipation Carnot-like refrigerator under $\chi$ and $\dot{\Omega}$ optimization criteria are obtained with power law dissipation level $\delta=1$. The comparative analysis of the extreme bounds of COP at optimized $\dot{{\Omega}}$ and $\chi$
figure of merit shows the lower bound of the $\dot{\Omega}$ optimized  COP is always higher than that of the $\chi$ optimized COP, while the upper bound of $\dot{\Omega}$ optimized COP is lesser than the $\chi$ optimized COP. 
\end{abstract}

\maketitle

\section{Introduction}
Finite time thermodynamic optimization to improve the performance of heat engines and refrigerators are attracting interest recent years due to the fact of providing more realistic theoretical bounds \cite{hernandez1, Hoffmann, chen, bejan, lchen}.  Refrigerator, a thermodynamical system that allows the transfer of heat from the source at a lower temperature $T_{c}$ to the heat sink at a higher temperature $T_{h}$. It is well known from the second law of thermodynamics that the heat cannot spontaneously flow from a region of lower temperature to a region of higher temperature \cite{callen}.  Hence, work is required to achieve this heat transfer process.  Refrigerator with a work input of $W$ completes the cycle of transfer of $Q_{c}$ amount of heat absorbed by the gas from a low temperature source and $Q_{h}$ amount of heat rejected to a high temperature heat sink. The Co-efficient Of Performance (COP) of the refrigerator operating between the two reservoirs is defined as, 
\begin{equation}
\epsilon = \frac{Q_{c}}{Q_{h}-Q_{c}}.
\label{int1}
\end{equation} 
This $\epsilon$ is bounded below the Carnot's co-efficient of performance,
\begin{equation}
\epsilon_{C} = \frac{T_{c}}{T_{h}-T_{c}}
\label{int2}
\end{equation} which requires infinite time to complete a cycle.  But in real scenario one can achieve maximum cooling rate at finite interval of time. By considering the irreversibilty of finite time heat transfer, Yvon \cite{yvon}, Novikov \cite{novikov}, Chambadal \cite{Chambadal} and later Curzon and Ahlborn \cite{curzon} extended the reversible Carnot cycle to an endoreversible Carnot cycle hence paving the growth of a new field called Finite time thermodynamics. These studies  provided more realistic limits for real engine/refrigerator performances under the finite time conditions, which ignited the search for universalities in performance of heat engines/refrigerators.

Different model systems were reported earlier to optimize the performance and to find the universal bounds to the efficiency and coefficient of performance of heat engines \cite{esposito, heatengine} and refrigerators \cite{velasco, zyan, detomas, refrigerator}, respectively. In particular, low-dissipation Carnot-like engines was investigated by Esposito et al.\cite{esposito}. Under the assumption that the irreversible entropy production in each isothermal process is inversely proportional to the time required for completing that process, they obtained the minimum and maximum bounds on the efficiency of the low dissipation engine. Where as per-unit time efficiency was proposed as a criterion to obtain bounds on the efficiency of heat engines by Ma \cite{ma}.  This per-unit time efficiency provided the compromise between the efficiency and speed of the thermodynamic cycle. However, A.C. Hernandez et.al., proved that {the endoreversible heat engine's efficiency at maximum per-unit-time efficiency is bounded between $\eta_{c}/2$ and $1 -\sqrt{1-\eta_{C}}$ \cite{hernandez}, where $\eta_{C}$ is the Carnot engine efficiency. Many phenomenological models of finite time heat engine were proposed to study the universal bounds on the efficiency at maximum power \cite{lowdiss}. Few studies  were also reported on Carnot like heat engine with non-adiabatic dissipation in finite time adiabatic processes, and showed that the additionally incorporated non-adiabatic dissipative term does not influence the extreme bounds on the efficiency at maximum power \cite{he, he2}. Apart from these, Cavina et.al., reported a microscopic model of quantum heat engine and obtained the universal nature of lower and upper bounds on the efficiency at maximum power of Carnot like heat engines with power law like dissipation \cite{cavina,yang}. Recently, one of the present author studied the efficiency at maximum power of  Carnot-like heat engines operate in a finite time under the power law dissipation regime and also showed that the generalized extreme bounds on the efficiency at maximum power does not influenced by the additionally incorporated non-adiabatic dissipative term \cite{pon2,pon}. It is also very clear from literature that, in the heat engine models, the power output and the per-unit-time efficiency are commonly used criterion of optimization though many other optimizations criteria \cite{hern} are used for better performance of heat engine.

Whereas in the case of refrigerators finding a suitable optimization criterion to determine its corresponding co-efficient of performance is very difficult \cite{hernandez1, hernandez}. However, numerous optimization criteria were also proposed to determine the refrigerator’s co-efficient of performance \cite{refrigerator}. For instance, by using per-unit-time co-efficient of performance as target function, Velasco et al., found the upper bound of the endoreversible refrigerators operating at the maximum per-unit-time co-efficient of performance to be $\epsilon_{CA}=\sqrt{1+\epsilon_{C}}-1$ \cite{velasco}. Yan and Y. Wang used $\chi = \epsilon Q_{c}/\tau$, as target function to optimize the performance of refrigerators and found the bounds of co-efficient of performance at maximum $\chi$. They reported that the $\chi$ figure of merit is the most appropriate criterion for the optimization of performance of refrigerators \cite{zyan}.  On the other hand, de Tomas et al., obtained the bounds on the co-efficient of performance under the symmetric low-dissipation condition using the optimization criterion $\dot{\Omega}=(2\epsilon-\epsilon_{max})W/\tau$. Here $\tau$ is the total time taken to complete the cyclic process and $\epsilon_{max}$ is the maximum co-efficient of performance \cite{detomas}. In most of the previous studies involving finite time thermodynamics of refrigerator, the non-adiabatic dissipation was not taken into account. Recently Y. Hu et.al., considered a Carnot like refrigerator with the non-adiabatic dissipation (the dissipation due to the effects of inner friction during the finite time adiabatic process) and analyzed the co-efficient of performance optimized with the $\chi$ and $\dot{\Omega}$ Criteria in a generalized setting with low dissipation assumption as a special case \cite{yhu}. Though there are studies on refrigerators considering low-dissipation and  non-adiabatic dissipation, there are seldom reports on the performance of refrigerator other than these dissipative regimes.

The heat engine model incorporating power law dissipation provides the generalized universal nature of extreme bounds on the efficiency at maximum power \cite{pon2, pon}.  Further,  the minimum and maximum bounds on the efficiency at the maximum power obtained in the  power law dissipative Carnot-like heat engines are unaffected by the non-adiabatic dissipation \cite{he, he2, pon}.  It is therefore very significant to consider a more generalized power law dissipative Carnot like refrigerator which also involving the non-adiabatic dissipation. Hence, in the present paper, a power law dissipative Carnot like refrigerator cycle of two irreversible isothermal and two irreversible adiabatic processes  with finite time non-adiabatic dissipation is considered and the co-efficient of performance under two optimization criteria $\chi$ and $\dot{\Omega}$ is studied. The generalized extreme bounds of the optimized co-efficient of performance under the above said optimization criteria are obtained. To the best of our knowledge,  this is an initial attempt to consider Carnot-like refrigerator operates in finite time under a different dissipation regime other than low-dissipation.

This paper is organized as follows: In section II, the model of  power law dissipative Carnot like refrigerator  is explained. In section III and IV, the optimization of co-efficient of performance at maximum "$\chi$" figure of merit and at maximum "$\dot{\Omega}$" figure of merit are derived and its extreme bounds are discussed. The paper concludes with the conclusion in section V.

\section{Power law dissipative Carnot-like Refrigerator}
A power law dissipative Carnot-like refrigerator is considered and it follows a cycle composed of two isotherms of finite time duration and two finite time adiabats with non-adiabatic dissipation \cite{yhu}.  During the isothermal expansion, the working substance is in contact with a cold reservoir at a constant temperature $T_{c}$, while during the isothermal compression, the working substance is in contact with hot heat reservoirs at constant temperature $T_{h}$. Let $t_{c}$ and $t_{h}$ denotes the finite time duration of the isothermal expansion and compression respectively. It is well known that in the ideal case, any adiabatic process is isentropic. Whereas in the present model the non-adiabatic dissipation is considered and hence the adiabatic process is nonisentropic.  The non-adiabatic dissipation develops additional heat which produces additional irreversible entropy production during the adiabatic process \cite{yhu}. Let $t_{a}$ and $t_{b}$ denotes the finite time duration of the adiabatic expansion and compression respectively. The details about all the four processes involved in the present model are discussed below:

\begin{itemize}
	\item Isothermal expansion:
	During this process, the working substance is in contact with the  cold reservoir at a lower temperature $T_{c}$ for a time interval $t_{c}$. In this process there is an exchange of $Q_{c}$ amount of heat between the working substance and the cold reservoir and 
	the variation of entropy is given as \cite{yhu}, 
	\begin{equation}
	\Delta S=\Delta S_c=Q_{c}/T_{c}+\Delta S^{ir}_{c}
	\label{isoexp}
	\end{equation}
	where $\Delta {S^{ir}_{c}}$ is the irreversible entropy production. 
	Here we used the convention that heat flow in to system is taken as positive.
	\item Adiabatic Expansion:
Due to the non-adiabatic dissipation, there is an increase in entropy during this adiabatic expansion process. The irreversible entropy production during the time interval $t_{c} < t < t_{c} + t_{a}$ is denoted by, 
\begin{equation}
\Delta S^{ir}_{a} = S_{a}- S_{c}
\label{adiabexp}
\end{equation}
where $S_{a}$ and  $S_{c}$ denotes the entropy at the instant $t_{c}+ t_{a}$ and $t_{c}$, respectively.
	\item Isothermal Compression:
Now the the working substance is in contact with the high temperature ($T_{h}$) hot reservoir for the time period $t_{c}+ t_{a} < t < t_{c} + t_{a}+t_{h}$ with the exchange of $Q_{h}$ amount of heat between the working substance and the hot reservoir.  The variation of entropy is given as,  
\begin{equation}
\Delta S_h=-Q_{h}/T_{h}+\Delta S^{ir}_{h}
\label{isocomp}
\end{equation}
 where $\Delta S^{ir}_{h} $ being the irreversible entropy production.   
	\item Adiabatic Compression:
	In the process of adiabatic compression during the time interval $t_{c}+ t_{a} +t_{h} < t < t_{c} + t_{a}+t_{h} +t_{b}$, the working substance is removed from the hot reservoir and now the entropy production due to non-adiabatic dissipation is given by,
\begin{equation}
\Delta S^{ir}_{b} = S_{b}- S_{h}
\label{adiabcomp}
\end{equation}
where $S_{b}$ and  $S_{h}$  denotes the entropy at the instant
$t_{c} + t_{a}+t_{h} +t_{b}$ and $t_{c}+ t_{a} +t_{h}$, respectively.

At the instance of completing the single cycle, the system recovers to its initial state and the total change in entropy of the system is zero, ie., $\Delta S+\Delta S^{ir}_{a}+\Delta S^{ir}_{b}+\Delta S_{h} = 0$ \cite{yhu}. Therefore, 
\begin{equation}
\Delta S_{h}=-(\Delta S+\Delta S^{ir}_{a}+\Delta S^{ir}_{b}).
\label{deltas}
\end{equation}
\end{itemize}

It is well known that there is a dependence of $1/\tau$ scaling ($\tau$ is the controlling time in which the process takes place) of the irreversible entropy production in all the four finite time processes(two isothermal and two adiabatic) \cite{esposito,ma}. But some recent studies showed that the irreversible entropy production in a finite time adiabatic process is associated with a $1/\tau^2$ scaling \cite{square}. This naturally leads to an assumption of various powers of $\tau$ to exactly match the dissipation in real heat engines \cite{pon, ponnil}. We believe strongly that a concept which is true for heat engines, can equally be valid for refrigerators. Thus a more generalized power law dissipative model describing all levels of dissipation of real refrigerators have been proposed and validated in the present work. Considering these facts and from Eqs.~(\ref{isoexp}) to ~(\ref{adiabcomp}), the irreversible entropy production associated with the isothermal and adiabatic processes can be written in a generalized power law dissipative form as \cite{pon2,pon}, 

\begin{equation}
\Delta S^{ir}_{i} = \alpha_{i}\left(\frac{\sigma_{i}}{t_{i}}\right)^{1/\delta}
\label{iso1}
\end{equation} 
with $i:a,b,c,h$ and $\sigma_{i} = \lambda_{i}\Sigma_{i}$ in which $\alpha_{i}$ and $\lambda_{i}$ are the
tuning parameters and $\Sigma_{i}$ are the dissipation coefficients for isothermal and adiabatic processes. The parameter, $\alpha_{i}$ provides the internal tuning of the system energy level and $\lambda_{i}$ provides external control to drive the system during adiabatic and isothermal processes \cite{he, he2}.  
The presence of $\delta \ge 0$ in the above expression signifies the level of dissipation present in the system.
 $\delta =1$ denotes that the system is in normal or low-dissipation regime, and $0<\delta <1$ and $\delta >1$ indicates that the system is in sub dissipation regime and super dissipation regime respectively \cite{yang,pon}.

Considering Eqs.~(\ref{isoexp}) to (\ref{iso1}), the amount of heat exchanged $Q_{c}$ and $Q_{h}$ can be obtained as follows \cite{pon,yhu}:
\begin{equation}
Q_{c} = T_{c}\left(\Delta S - \alpha_{c}\left(\frac{\sigma_{c}}{t_{c}}\right)^{1/\delta}\right)
\label{qc}
\end{equation}
 
\begin{equation}
Q_{h} = T_{h}\left(\Delta S + \sum_{i=a,b,h}\alpha_{i}\left(\frac{\sigma_{i}}{t_{i}}\right)^{1/\delta}\right).
\label{qh}
\end{equation}
 Using Eqs.~(\ref{qc}) and ~(\ref{qh}), the following relation can be found:
\begin{equation}
\frac{Q_{h}}{T_{h}}-\frac{Q_{c}}{T_{c}} = \sum_{i=c,a,b,h} \alpha_{i} \left(\frac{\sigma_{i}}{t_{i}}\right)^{1/\delta}.
\label{qhqc}
\end{equation}
The work input to the refrigerator to complete the cycle of transfer of $Q_{c}$ amount of heat absorbed by the working substance from a low temperature source and $Q_{h}$ amount of heat rejected to a high temperature heat sink in the total time period $ t=t_{c}+t_{a}+t_{b}+t_{h} $ is given by \cite{yhu},
\begin{equation}
W = Q_{h}-Q{c}.
\label{work}
\end{equation}
Using Eqs.~(\ref{qc}) and ~(\ref{qh}), the expression for work done can be written as,

\begin{equation}
W = (T_{h}-T_{c})\Delta S +T_{c} \alpha_{c}\left(\frac{\sigma_{c}}{t_{c}}\right)^{1/\delta}+T_{h}\sum_{i=a,b,h}\alpha_{i}\left(\frac{\sigma_{i}}{t_{i}}\right)^{1/\delta}.
\label{w1}
\end{equation}
The co-efficient of performance of the refrigerator (Eq.\ref{int1}) is then given by, 
\begin{equation}
\epsilon = \frac{T_{c}\left(\Delta S - \alpha_{c}\left(\frac{\sigma_{c}}{t_{c}}\right)^{1/\delta}\right)}{(T_{h}-T_{c})\Delta S +T_{c}\alpha_{c}\left(\frac{\sigma_{c}}{t_{c}}\right)^{1/\delta}+T_{h}\sum_{i=a,b,h}\alpha_{i}\left(\frac{\sigma_{i}}{t_{i}}\right)^{1/\delta}}.
\label{epsilon}
\end{equation}

In the present work,  $\chi$ figure of merit and the $\dot{\Omega}$ figure of merit are optimized for analyzing the performance of refrigerator with (isothermal and non-adiabatic) power law dissipation.  

\section{Co-efficient of performance at maximum "$\chi$" figure of merit}
The $\chi$ figure of merit is defined as the product of co-efficient of performance of a refrigerator $\epsilon$ times the heat exchanged between the working substance and the cold reservoir $Q_{c}$, per the total time duration to complete a cycle $t$, 
$\chi= \epsilon Q_{c}/t$. The "$\chi$" figure of merit can be obtained by substituting equation~(\ref{qc}) and ~(\ref{epsilon}) in $\chi= \epsilon Q_{c}/t$ as, 
\begin{widetext}
\begin{equation}
\chi = \frac{T_{c}^{2}\left(\Delta S - \alpha_{c}\left(\frac{\sigma_{c}}{t_{c}}\right)^{1/\delta}\right)^{2}}{t \left[(T_{h}-T_{c})\Delta S +T_{c}\alpha_{c}\left(\frac{\sigma_{c}}{t_{c}}\right)^{1/\delta}+T_{h}\sum_{i=a,b,h}\alpha_{i}\left(\frac{\sigma_{i}}{t_{i}}\right)^{1/\delta}\right]}
\label{chi}
\end{equation}
where $t=t_{c}+t_{a}+t_{b}+t_{h}$.
Optimizing the "$\chi$" figure of merit with respect to time $t_{i}(i:c,h,a,b)$ gives the values of $\tilde{t_{i}}(i:c,h,a,b)$ at which "$\chi$" is maximum. The values for $\tilde{t_{i}}(i:c,h,a,b)$ by considering $\frac{\partial \chi}{\partial t_{i}} = 0$ are given below:
\begin{equation}
\tilde{t_{a}} = \left[\Phi_a \left\{\left(\frac{1}{\delta}-\frac{\epsilon}{2+\epsilon}\right)\left(\frac{T_{c}\alpha_{c}\sigma_{c}^{1/\delta}}{T_{h}\alpha_{a}\sigma_{a}^{1/\delta}}\left(\frac{2+\epsilon}{\epsilon}\right)\right)^{\frac{\delta}{\delta+1}}+\left(\frac{1}{\delta}-1\right)\left(1+\sum_{i=b,h}\left(\frac{\alpha_{i}\sigma_{i}^{1/\delta}}{\alpha_{a}\sigma_{a}^{1/\delta}}\right)^{\frac{\delta}{\delta+1}}\right)\right\}\right]^{\delta},
\label{ta}
\end{equation}
\begin{equation}
\tilde{t_{b}} = \left[\Phi_b \left\{\left(\frac{1}{\delta}-\frac{\epsilon}{2+\epsilon}\right)\left(\frac{T_{c}\alpha_{c}\sigma_{c}^{1/\delta}}{T_{h}\alpha_{b}\sigma_{b}^{1/\delta}}\left(\frac{2+\epsilon}{\epsilon}\right)\right)^{\frac{\delta}{\delta+1}}+\left(\frac{1}{\delta}-1\right)\left(1+\sum_{i=a,h}\left(\frac{\alpha_{i}\sigma_{i}^{1/\delta}}{\alpha_{b}\sigma_{b}^{1/\delta}}\right)^{\frac{\delta}{\delta+1}}\right)\right\}\right]^{\delta},
\label{tb}
\end{equation}
\begin{equation}
\tilde{t_{h}} = \left[\Phi_h \left\{\left(\frac{1}{\delta}-\frac{\epsilon}{2+\epsilon}\right)\left(\frac{T_{c}\alpha_{c}\sigma_{c}^{1/\delta}}{T_{h}\alpha_{h}\sigma_{h}^{1/\delta}}\left(\frac{2+\epsilon}{\epsilon}\right)\right)^{\frac{\delta}{\delta+1}}+\left(\frac{1}{\delta}-1\right)\left(1+\sum_{i=a,b}\left(\frac{\alpha_{i}\sigma_{i}^{1/\delta}}{\alpha_{h}\sigma_{h}^{1/\delta}}\right)^{\frac{\delta}{\delta+1}}\right)\right\}\right]^{\delta},
\label{th}
\end{equation}
\begin{equation}
\tilde{t_{c}}= \left[\Phi_c \left\{\left(\frac{2+\epsilon}{\epsilon}\right)\left(\frac{1}{\delta}-1\right)\left(\sum_{i=a,b,h}\frac{T_{h}\alpha_{i}\sigma_{i}^{1/\delta}}{T_{c}\alpha_{c}\sigma_{c}^{1/\delta}}\left(\frac{\epsilon}{2+\epsilon}\right)\right)^{\frac{\delta}{\delta+1}} +\left(\frac{2+\epsilon}{\epsilon}\right)-1\right\}\right]^{\delta},
\label{tc}
\end{equation}
\end{widetext}
where $\Phi_j=\frac{T_{h}\alpha_{j}\sigma_{j}^{1/\delta}}{(T_{h}-T_{c})\Delta S}$ $(j=a,b,h)$ and
$\Phi_c=\frac{T_{c}\alpha_{c}\sigma_{c}^{1/\delta}}{(T_{h}-T_{c})\Delta S}$.
It should be noted that the above expressions (Eqs. \ref{ta}-\ref{tc}) contain $\epsilon$ in the right
 hand side and hence tedious and cumbersome to simply further. However, one can observe that the above expressions are sufficient for further detailed analysis. 
Considering $\frac{\partial\chi}{\partial t_{i}}=0,(i=a,b,c,h)$, four following relations for $\chi$ can also be obtained.
\begin{equation}
\chi_{\left(\frac{\partial\chi}{\partial t_{i}}=0\right)} =\frac{\epsilon^{2}T_{h}\alpha_{i}\sigma_{i}^{\frac{1}{\delta}}}{\delta \tilde{t_{i}}^{\frac{1}{\delta}+1}}
\label{chii}
\end{equation}
where in Eq.~(\ref{chii}), $i=a,b,h$. Similarly, the value of $\chi$, when $\frac{\partial\chi}{\partial t_{c}}=0$ is,
\begin{equation}
\chi_{\left(\frac{\partial\chi}{\partial t_{c}}=0\right)} = \frac{\epsilon T_{c}\alpha_{c}\sigma_{c}^{\frac{1}{\delta}}\left(2+\epsilon\right)}{\delta \tilde{t_{c}}^{\frac{1}{\delta}+1}}.
\label{chic}
\end{equation}
The ratios of $\frac{\tilde{t_{c}}}{\tilde{t_{i}}}(i: a,b,h)$ can also be obtained from the optimized $"\chi"$ and are given below:
\begin{equation}
\left(\frac{\tilde{t_{c}}}{\tilde{t_{i}}}\right)^{\frac{1}{\delta}+1} = \frac{T_{c}\alpha_{c}\sigma_{c}^{1/\delta}}{T_{h}\alpha_{i}\sigma_{i}^{1/\delta}}\left(\frac{2+\epsilon}{\epsilon}\right).
\label{tcti}
\end{equation}
Similarly the ratios for $\tilde{t_{j}}/\tilde{t_{i}}$ with $i,j=a,b,h$ are also given by,

\begin{equation}
\left(\frac{\tilde{t_{j}}}{\tilde{t_{i}}}\right)^{\frac{1}{\delta}+1} = \frac{\alpha_{j}\sigma_{j}}{\alpha_{i}\sigma_{i}}.
\label{tjti}
\end{equation} 
From Eq.~(\ref{w1}),  
\begin{equation}
Q_{h}-Q_{c} = Q_{c}\left(\frac{T_{h}}{T_{c}}-1\right)+T_{h}\alpha_{c}\left(\frac{\sigma_{c}}{t_{c}}\right)^{1/\delta}+T_{h}\sum_{i=a,b,h}\alpha_{i}\left(\frac{\sigma_{i}}{t_{i}}\right)^{1/\delta}
\label{diffqhqc}
\end{equation}
Which yields,
\begin{equation}
\frac{Q_{h}-Q_{c}}{Q_{c}} = \frac{1}{\epsilon_{C}}+\frac{T_{h}\alpha_{c}}{Q_{c}}\left(\frac{\sigma_{c}}{t_{c}}\right)^{1/\delta}+\frac{T_{h}}{Q_{c}}\sum_{i=a,b,h}\alpha_{i}\left(\frac{\sigma_{i}}{t_{i}}\right)^{1/\delta}
\label{diff2}
\end{equation}
On further simplification using Eqs.(\ref{ta}-\ref{tc}), Eq.(\ref{tcti}) and Eq.(\ref{tjti}) with $\epsilon=\epsilon_{\chi}$, the above equation reduces to, 
\begin{equation}
\frac{1}{\epsilon_{\chi}} = \frac{1}{\epsilon_{C}}+ \frac{T_{h}}{T_{c}\left(\frac{\Delta S t_{c}^{1/\delta}}{\alpha_{c}\sigma_{c}^{1/\delta}-1}\right)} \left(1+\sum_{i=a,b,h}\frac{\alpha_{i}}{\alpha_{c}} \left(\frac{\sigma_{i}t_{c}}{\sigma_{c}t_{i}}\right)^{1/\delta}\right),
\label{diff3}
\end{equation}
where $\epsilon_{\chi}$ is the the co-efficient of performance at maximum $\chi$. Using Eqs.~(\ref{ta}) -~(\ref{tc}) and  ratios of $t_{i}$'s, the co-efficient of performance at maximum $\chi$ figure of merit can be obtained as follows:\\
\begin{equation}
\frac{1}{\epsilon_{\chi}} = \frac{1}{\epsilon_{C}}+ \frac{T_{h}}{T_{c}}\left\{\frac{1+\frac{T_{c}}{T_{h}} \Upsilon}{\epsilon_{C}\left[ \Upsilon \left(\frac{1}{\delta}-1\right)+\left(\frac{2+\epsilon_{\chi}}{\epsilon_{\chi}}\right)\frac{1}{\delta}-1\right]-1}\right\},
\label{diff4}
\end{equation}
where $\Upsilon=\left(\frac{2+\epsilon_{\chi}}{\epsilon_{\chi}}\right) \left(\frac{T_{h}}{T_{c}}\sum_{i=a,b,h}\frac{\alpha_{i}\sigma_{i}^{1/\delta}}{\alpha_{c}\sigma_{c}^{1\delta}}\left(\frac{\epsilon_{\chi}}{2+\epsilon_{\chi}}\right)\right)^{\frac{\delta}{\delta+1}}$.
Directly adding both sides of Eqs.~(\ref{chii}) \& ~(\ref{chic}) and using Eqs.~(\ref{qc}), ~(\ref{qh}) and ~(\ref{qhqc}), the final expression for Co-efficient of performance at maximum $\chi$ figure of merit can be obtained as, 
\begin{equation}
\frac{1}{\epsilon_{\chi}}-\frac{1}{\epsilon_{C}} = \frac{4\delta}{t}\left(\frac{1}{\varsigma_{1}\epsilon_{\chi}+\varsigma_{2}\left(\frac{2\epsilon_{C}-\epsilon_{\chi}}{1+\epsilon_{C}}\right)}\right).
\label{diff5}
\end{equation} 
Solving the above equation, we get
\begin{equation}
\epsilon_{\chi} = \frac{\epsilon_{C}\left\{\varphi+\sqrt{\varphi^{2}+8\varsigma_{2}t^{2}((\varsigma_{1}-\varsigma_{2})+\varsigma_{1}\epsilon_{C})}\right\}}{2t(\varsigma_{1}-\varsigma_{2})+\varsigma_{1}\epsilon_{C}}
\label{finale}
\end{equation}
where $\varphi=\left(t\varsigma_{1}-4\delta-3t\varsigma_{2}+(t\varsigma_{1}-4\delta)\epsilon_{C}\right)$,
\begin{equation*}
\varsigma_{1}= \frac{\sum_{i=a,b,c,h}\frac{\alpha_{i}\sigma_{i}^{1/\delta}}{t_{i}^{\frac{1}{\delta}+1}}}{\sum_{i=a,b,c,h}\frac{\alpha_{i}\sigma_{i}^{1/\delta}}{t_{i}^{\frac{1}{\delta}}}}
\end{equation*}
and 
\begin{equation*}
\varsigma_{2}= \frac{\frac{\alpha_{c}\sigma_{c}^{1/\delta}}{t_{c}^{\frac{1}{\delta}+1}}}{\sum_{i=a,b,c,h}\frac{\alpha_{i}\sigma_{i}^{1/\delta}}{t_{i}^{\frac{1}{\delta}}}}.
\end{equation*}
Neglecting the adiabatic dissipation co-efficients, $\sigma_{a}=0$ and $\sigma_{b}=0$, the co-efficient of performance as derived in equation~(\ref{diff5}) with $\delta=1$  reduces to the one derived for Carnot-like refrigerators without adiabatic dissipation by Y Wang et.al.,\cite{hernandez1}.

It can be observed from the Eq.~(\ref{diff4}), the value of co-efficient of performance at the maximum $\chi$ figure of merit depends on the ratio between values of $\sigma_{i}'s$ and $\sigma_{c}$. The generalized extreme bounds of the co-efficient of performance at maximum "$\chi$" figure of merit are obtained from Eq.(\ref{diff4}) as 
\begin{equation}
\epsilon_{C}(1-\delta)\equiv\epsilon^{-}_{\chi}\leq\epsilon_{\chi}\leq\epsilon^{+}\equiv \frac{\left(\zeta+\sqrt{\zeta^2+8\epsilon_{C}}\right)}{2}
\label{bound}
\end{equation} 
where, $\zeta = \epsilon_{C}(1-\delta)-(\delta+2)$.  These extreme lower and upper bounds of the co-efficient of performance at maximum "$\chi$" figure of merit are achieved when $\sigma_{c}\rightarrow 0$ and $\sigma_{c}\rightarrow \infty$, respectively.
When $\delta=1$, the lower bound becomes $0$ for $\sigma_{c}\rightarrow 0$ and the upper bound becomes $(\sqrt{9+8\epsilon_{C}}-3)/2$ for $\sigma_{c}\rightarrow \infty$, which is the bound of the co-efficient of performance at the maximum $\chi$ figure of merit obtained for low dissipation case \cite{yhu}. Since $\epsilon_{\chi}$ cannot be negative the lower bound can be rewritten as $\epsilon^{-}_{\chi} = \frac{1}{2}\epsilon_{C}(\mid 1-\delta \mid+(1-\delta))$ which is equal to $(1-\delta)\epsilon_{C}$ when $\delta < 1$ and  $0$ when $\delta \ge 1$ and the upper bound remains positive for all values of $\delta$. Thus, a more generalized upper and lower bounds on the co-efficient of performance can be obtained under the combined adiabatic and isothermal power law dissipation in the asymmetric limits.

\section{Co-efficient of performance at maximum "$\dot{\Omega}$" figure of merit}

This section discusses the optimization of $\dot{\Omega}$ figure of merit and its significance in detail.
The $\dot{\Omega}$ figure of merit is defined as the product of difference between twice the co-efficient of performance of a refrigerator $\epsilon$ and maximum co-efficient of performance of a refrigerator $\epsilon_{max}$ and the work required by the system $W$, divided by the total time duration required to complete a single cycle $t$, $\dot{\Omega}=(2\epsilon-\epsilon_{max})W/t$.
The $\dot{\Omega}$ can be expressed using Eq.(\ref{w1}) and Eq.(\ref{diffqhqc}) with $\epsilon_{max}=\epsilon_{C}$ as,
\begin{equation}
\dot{\Omega} = \frac{1}{t}\left\{2\left(T_{c}\Delta S -T_{c}\alpha_{c}\left(\frac{\sigma_{c}}{t_{c}}\right)^{\frac{1}{\delta}}\right)-\epsilon_{C} \it{W} \right\},
\label{omega}
\end{equation}
where 
\begin{equation*}
\it{W}=\left[(T_{h}-T_{c})\Delta S +T_{c}\alpha_{c}\left(\frac{\sigma_{c}}{t_{c}}\right)^{\frac{1}{\delta}}+T_{h}\sum_{i=a,b,h}\alpha_{i}\left(\frac{\sigma_{i}}{t_{i}}\right)^{\frac{1}{\delta}}\right].
\end{equation*}
Similar to $\chi$ figure of merit, the Optimizing the "$\dot{\Omega}$" figure of merit with respect to the time $t_{i}(i:c,h,a,b)$ gives the values of $\tilde{t_{i}}(i:c,h,a,b)$ at which "$\dot{\Omega}$" is maximum. The values for $\tilde{t_{i}}(i:c,h,a,b)$ by considering $\frac{\partial \dot{\Omega}}{\partial t_{i}} = 0$ are given below:
\begin{widetext}
\begin{equation}
\tilde{t_{a}} = \left\{\frac{T_{h}\alpha_{a}\sigma_{a}^{\frac{1}{\delta}}}{\Delta S(T_{h}-T_{c})}\left[\left(1+\frac{1}{\delta}\right)\left[1+\left(\left(\frac{\epsilon_{C}+2}{\epsilon_{C}+1}\right)\left(\frac{\alpha_{c}\sigma_{c}^{\frac{1}{\delta}}}{\alpha_{a}\sigma_{a}^{\frac{1}{\delta}}}\right)\right)^{\frac{\delta}{\delta+1}}\right]+\sum_{i=b,h}\left(\frac{\alpha_{i}\sigma_{i}^{\frac{1}{\delta}}}{\alpha_{a}\sigma_{a}^{\frac{1}{\delta}}}\right)^{\frac{\delta}{\delta+1}}\right]\right\}^{\delta},
\label{taomega}
\end{equation}
\begin{equation}
\tilde{t_{b}} = \left\{\frac{T_{h}\alpha_{b}\sigma_{b}^{\frac{1}{\delta}}}{\Delta S(T_{h}-T_{c})}\left[\left(1+\frac{1}{\delta}\right)\left[1+\left(\left(\frac{\epsilon_{C}+2}{\epsilon_{C}+1}\right)\left(\frac{\alpha_{c}\sigma_{c}^{\frac{1}{\delta}}}{\alpha_{b}\sigma_{b}^{\frac{1}{\delta}}}\right)\right)^{\frac{\delta}{\delta+1}}\right]+\sum_{i=a,h}\left(\frac{\alpha_{i}\sigma_{i}^{\frac{1}{\delta}}}{\alpha_{b}\sigma_{b}^{\frac{1}{\delta}}}\right)^{\frac{\delta}{\delta+1}}\right]\right\}^{\delta},
\label{tbomega}
\end{equation}
\begin{equation}
\tilde{t_{h}} = \left\{\frac{T_{h}\alpha_{h}\sigma_{h}^{\frac{1}{\delta}}}{\Delta S(T_{h}-T_{c})}\left[\left(1+\frac{1}{\delta}\right)\left[1+\left(\left(\frac{\epsilon_{C}+2}{\epsilon_{C}+1}\right)\left(\frac{\alpha_{c}\sigma_{c}^{\frac{1}{\delta}}}{\alpha_{h}\sigma_{h}^{\frac{1}{\delta}}}\right)\right)^{\frac{\delta}{\delta+1}}\right]+\sum_{i=a,b}\left(\frac{\alpha_{i}\sigma_{i}^{\frac{1}{\delta}}}{\alpha_{h}\sigma_{h}^{\frac{1}{\delta}}}\right)^{\frac{\delta}{\delta+1}}\right]\right\}^{\delta},
\label{thomega}
\end{equation}
\begin{equation}
\tilde{t_{c}}=\left\{\frac{\alpha_{c}\sigma_{c}^{\frac{1}{\delta}}}{\Delta S}\left[\left(1+\frac{1}{\delta}\right)(2+\epsilon_{C}\left[1+\left(\frac{\epsilon_{C}+1}{\epsilon_{C}+2}\right)^{\frac{\delta}{\delta+1}}\sum_{i=a,b,h}\left(\frac{\alpha_{i}\sigma_{i}^{\frac{1}{\delta}}}{\alpha_{c}\sigma_{c}^{\frac{1}{\delta}}}\right)^{\frac{\delta}{\delta+1}}\right]\right]\right\}^{\delta}.
\label{tcomega}
\end{equation}
\end{widetext}
The ratios of $\frac{\tilde{t_{c}}}{\tilde{t_{i}}}(i: a,b,h)$ can also be obtained from the optimized $"\dot{\Omega}"$ and are given below:
\begin{equation}
\left(\frac{\tilde{t_{c}}}{\tilde{t_{i}}}\right)^{\frac{1}{\delta}+1} = \left(\frac{\epsilon_{C}+1}{\epsilon_{C}+2}\right)\left(\frac{\alpha_{i}\sigma_{i}^{1/\delta}}{\alpha_{c}\sigma_{c}^{1/\delta}}\right).
\label{omegatcratio}
\end{equation}
Similarly the ratios for $\tilde{t_{j}}/\tilde{t_{i}}$ with $i,j=a,b,h$ are also given by,
\begin{equation}
\left(\frac{\tilde{t_{i}}}{\tilde{t_{j}}}\right)^{\frac{1}{\delta}+1} = \frac{\alpha_{i}\sigma_{i}^{1/\delta}}{\alpha_{j}\sigma_{j}^{1/\delta}}.
\label{omegatiratio}
\end{equation}
Substituting Eq.~(\ref{omegatcratio}) in Eq.~(\ref{diff3}), the co-efficient of performance at maximum $\dot{\Omega}$ figure of merit can be obtained as follows:
\begin{equation}
\epsilon_{\dot{\Omega}} = \frac{\epsilon_{C}\beta_{1}\left(1-\frac{1}{\beta_{1}}\right)}{\beta_{1}+\epsilon_{C}+\frac{\beta_{1}\beta_{2}}{\left(1+\frac{1}{\delta}\right)(1+\beta_{2})}}.
\label{epsidotomega}
\end{equation}
where, in the above equation, $\beta_{1}=(1+\frac{1}{\delta})(\epsilon_{C}+2)(1+\beta_{2})$ in which, 
\begin{equation}
\beta_{2}=\left(\frac{\epsilon_{C}+1}{\epsilon_{C}+2}\right)^{\frac{\delta}{\delta+1}}\sum_{i=a,b,h}\left(\frac{\alpha_{i}\sigma_{i}^{1/\delta}}{\alpha_{c}\sigma_{c}^{1/\delta}}\right)^{\frac{\delta}{\delta+1}}.
\label{eqv}
\end{equation}
It can also be observed from the Eq.~(\ref{epsidotomega}), the value of co-efficient of performance at the maximum $\dot{\Omega}$ figure of merit depends on the ratio between values of $\sigma_{i}'s$ and $\sigma_{c}$.  The extreme bounds of the co-efficient of performance at the maximum $\dot{\Omega}$ figure of merit is obtained when $\sigma_{c}\rightarrow 0$ and $\sigma_{c}\rightarrow \infty$.  That is, when $\sigma_{c}\rightarrow 0$, $\beta_{2}\rightarrow\infty$ for which $\epsilon_{\dot{\Omega}}=\frac{\delta+1}{2\delta+1}\epsilon_{C}$ and when $\sigma_{c}\rightarrow \infty$, $\beta_{2}\rightarrow 0$ for which $\epsilon_{\dot{\Omega}}=\frac{(\delta+2)+(\delta+1)\epsilon_{C}}{2(\delta+1)+(2\delta+1)\epsilon_{C}}\epsilon_{C}$. This shows that co-efficient of performance at the maximum $\dot{\Omega}$ figure of merit lies between these two extreme bounds, which is given by,
\begin{equation}
\frac{\delta+1}{2\delta+1} \epsilon_{C} \equiv\epsilon^{-}_{\dot{\Omega}} \leq \epsilon_{\dot{\Omega}} \leq \epsilon^{+}_{\dot{\Omega}} \equiv \frac{(\delta+2)+(\delta+1)\epsilon_{C}}{2(\delta+1)+(2\delta+1)\epsilon_{C}}\epsilon_{C}.
\label{boundomega}
\end{equation}
The generalized lower and upper bounds are obtained for the asymmetric dissipation limits of $\sigma_{c}\rightarrow 0$ and $\sigma_{c}\rightarrow \infty$, respectively, for any finite values of $\sigma_{i}, (i: a, b, h)$.  When $\delta=1$, the values of optimized co-efficient of performance at the maximum $\dot{\Omega}$ figure of merit of low dissipation regime is obtained, which is $\epsilon^{-}_{\dot{\Omega}} = \frac{2}{3}\epsilon_{C}$ and $\epsilon^{+}_{\dot{\Omega}}=\frac{3+2\epsilon_{C}}{4+3\epsilon_{C}}\epsilon_{C}$, the lower and upper bound respectively \cite{yhu}. Thus, the generalized universal nature of lower and upper bounds on the co-efficient of performance at maximum $\dot{\Omega}$ figure of merit (Eq.~(\ref{boundomega})) under the combinations of isothermal and adiabatic asymmetric dissipation limits is obtained.
 
\begin{widetext}
\section{Discussion}
The comparison is made between our predicted co-efficient of performance and the observed co-efficient of performance of some real refrigerators and is shown in Figures~\ref{fig1} and ~\ref{fig2}. For an illustrative purpose, we have taken only limited number of experimental datas \cite{refgbook} in our analysis.  Figure~\ref{fig1} shows the  lower and upper bounds of $\dot{\Omega}$ optimized co-efficient of performance plotted versus $\epsilon_C$ in the sub dissipation regime $0<\delta<1$, low dissipation regime $\delta=1$ and the super dissipation regime $\delta>1$. From the inset figure, one can hardly observe the appreciable difference between the lower and upper bounds of $\dot{\Omega}$ optimized co-efficient of performance for broader range of $\epsilon$. If we notice further in the inset of the figure that all the data points of real refrigerators  are found to lie within the lower and upper bounds of the $\dot{\Omega}$ figure of merit in the dissipation range $0.5 \le \delta \le 3$. This result validates the proposed model of power law dissipation to estimate the experimental results of real refrigerators working in the different dissipation regime.

\begin{figure*}[h]
\includegraphics[scale=0.6]{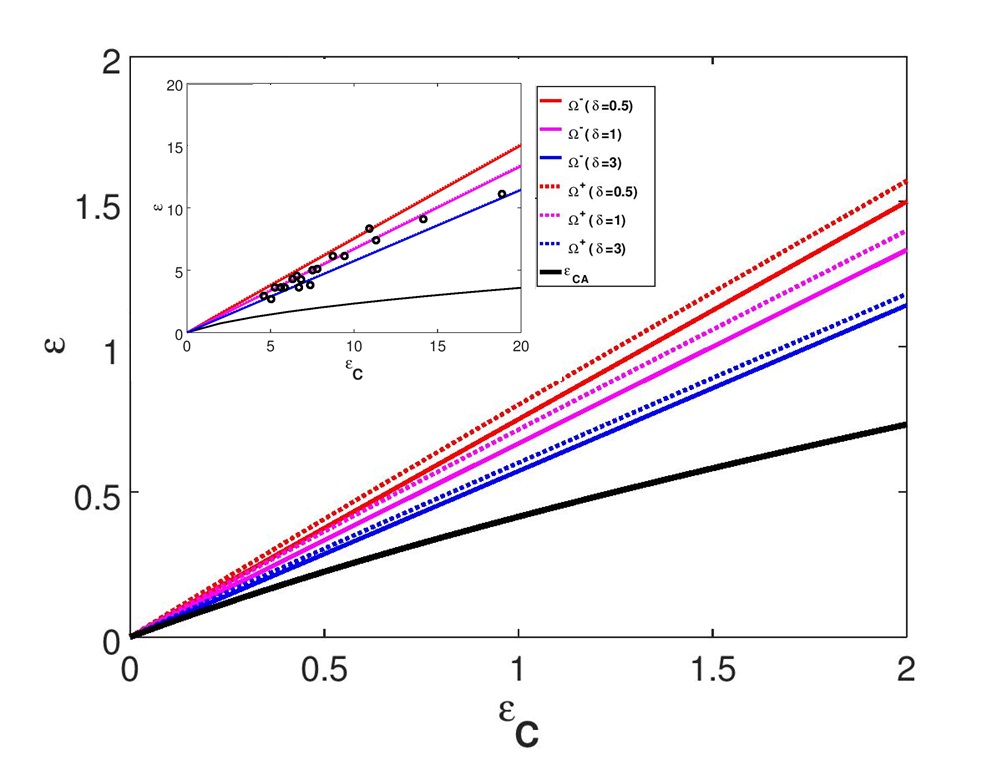}
\caption{The  lower (line) and upper bounds (dotted line) of $\dot{\Omega}$ optimized co-efficient of performance plotted versus $\epsilon_C$ in (top) the sub dissipation regime $\delta=0.5$, (mid) the low dissipation regime $\delta=1$ and (bot) the super dissipation regime $\delta=3$. Comparison between theoretical prediction optimized under maximum $\dot{\Omega}$ figure of merit and the observed co-efficient of performance of real refrigerators  is shown in the inset figure. This figure shows the experimental data (open circle) fitted well within the  results of the present model in the dissipation range $0.5 \le \delta \le 3$. In the inset figure, one can hardly notice the  appreciable difference between the lower and upper bounds of $\dot{\Omega}$ optimized co-efficient of performance for broader range of 
$\epsilon$. Curzon-Ahlborn co-efficient of performance $\epsilon_{CA}$ (lowest curve) is also shown in this plot for comparison.}
\label{fig1}
\end{figure*}

We have made the similar kind of analysis for $\chi$ figure of merit and is given in Figure~\ref{fig2}.  
Figure~\ref{fig2}-a, b and c shows the  lower and upper bounds of $\chi$ optimized  co-efficient of performance in the sub dissipation regime $0<\delta<1$, low dissipation regime $\delta=1$ and the super dissipation regime $\delta>1$. On the contrary to $\dot{\Omega}$ figure of merit, we could observe from this figure that all the data of real refrigerators located well below the upper bounds of all dissipation levels. However, the lower bound covers the experimental data only in the sub dissipation regime $0<\delta<1$ which is shown  
in Figure~\ref{fig2}-d for two different values of $\delta$, say $\delta=0.25$ and $\delta=0.5$. This result again validates the proposed model of power law dissipation to estimate the experimental results of real refrigerators working in different levels of dissipation.   

 \emph{
\begin{figure*}[h]
	\centering
		\includegraphics[scale=0.6]{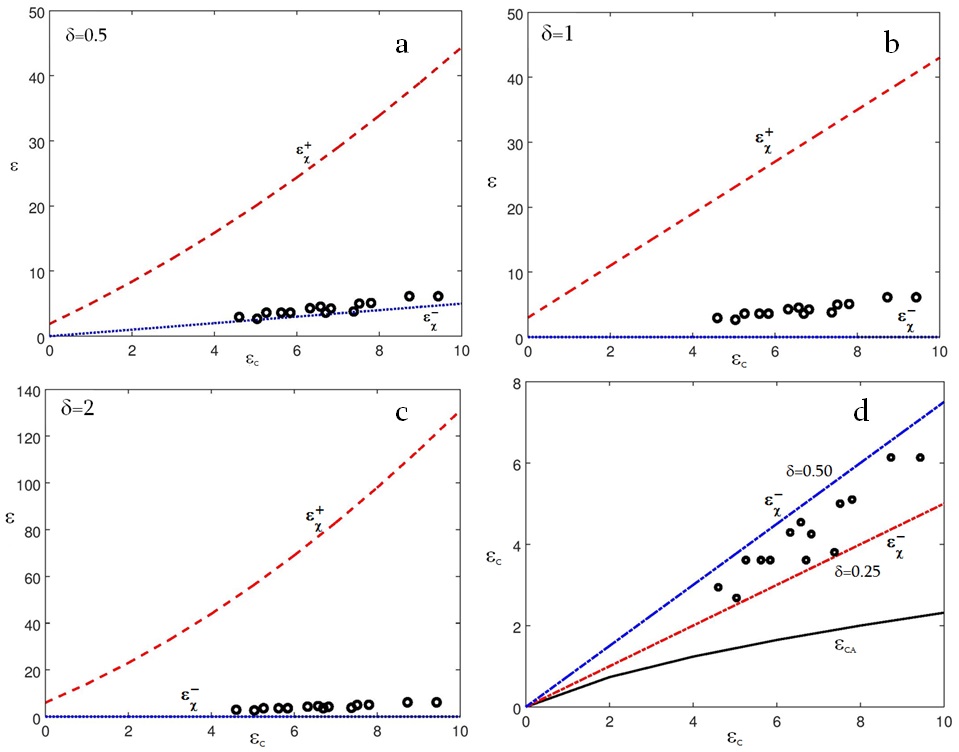}
		\caption{The  lower (dotted line) and upper bounds (dashed line) of $\chi$ optimized co-efficient of performance plotted versus $\epsilon_C$ in (a) the sub dissipation regime $\delta=0.5$, (b) the low dissipation regime $\delta=1$ and (c) the super dissipation regime $\delta=2$. Comparison between theoretical prediction optimized under maximum $\chi$ figure of merit and the observed co-efficient of performance of real refrigerators shows the experimental data (open circle) lies well below the upper bound of different dissipation levels. d) The lower bound encompasses the experimental data in the sub dissipation regime $0<\delta<1$ for two different values of 
$\delta=0.25$ and $\delta=0.5$. Curzon-Ahlborn co-efficient of performance $\epsilon_{CA}$ (lowest curve) is also shown in this figure
for comparison.}		
\label{fig2}
\end{figure*}}

We also compared the upper and lower bounds of co-efficient of performance at maximum $\dot{{\Omega}}$ and $\chi$ figure of merit. 
We found that, lower bound of the co-efficient of performance optimized  under maximum $\dot{\Omega}$ figure of merit is higher than that of the co-efficient of performance optimized under maximum $\chi$ figure of merit. But while comparing the upper bound, we could observe an interesting result that the upper bound  $\epsilon^{+}_{\chi}$ is more than the $\epsilon^{+}_{\dot{\Omega}}$. 
For an illustration, this is shown in Figure~\ref{fig3} for $\delta=0.5$. These results suggests that $\chi$ figure of merit seems to be a theoretically more valuable figure of merit but when comparing experimental datas, the ${\dot{\Omega}}$ figure of merit is more likely to fit in all the dissipation regimes.  Hence a detailed study with a huge data is required to analyze which figure of merit is best to fit the experimental data. Our analysis showed, the power law dissipation model can be regarded as a generalized model to fit any kind of dissipation in the real refrigerators.

\begin{figure*}[h]
	\includegraphics[scale=0.6]{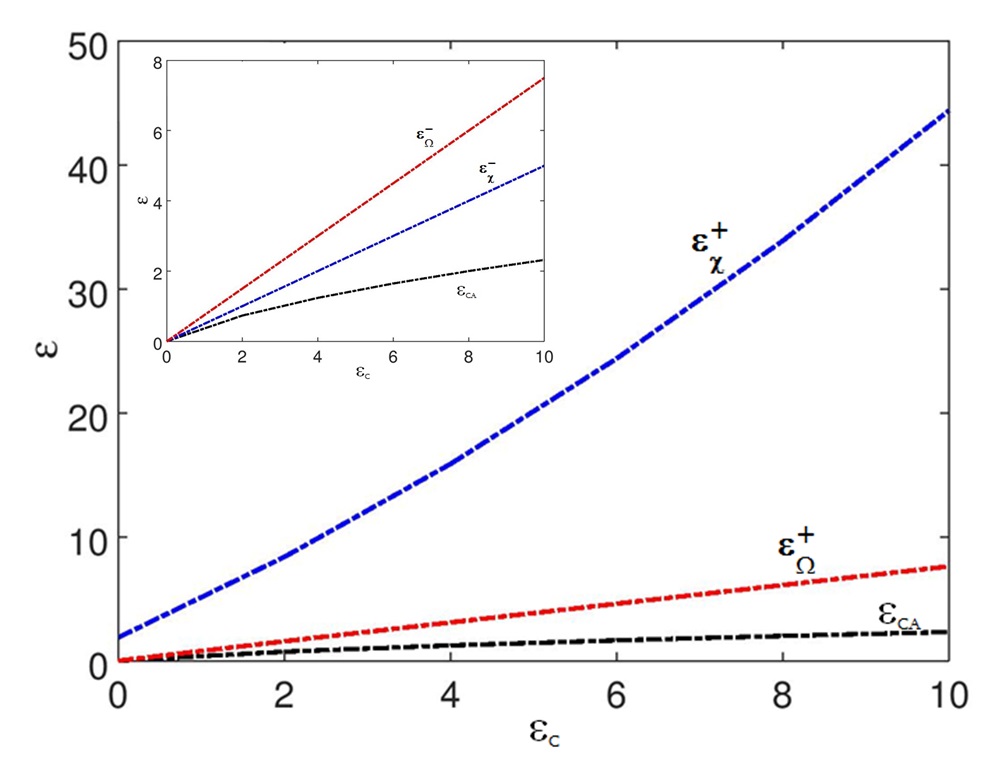}
	\caption{Figure showing comparison between the upper bounds of co-efficient of performance under maximum $\dot{\Omega}$ of merit and $\chi$ figure of merit for $\delta = 0.5$. Inset shows the same for the lower bound. $\epsilon_{CA}$ is the Curzon-Ahlborn co-efficient of performance.}
\label{fig3}
\end{figure*}

\end{widetext}}

\section{Conclusion}
In this paper, the  generalized  extreme bounds of the coefficient of performance for the power law dissipative Carnot-like refrigerator under $\chi$ and $\dot{\Omega}$ optimization criteria was investigated.  When $\delta=1$, the bounds of the co-efficient of performance  with the $\chi$ and $\dot{\Omega}$ figure of merit in the asymmetric dissipation converges to the same bounds as the corresponding ones obtained from previous low dissipation model. These results also showed that the presence of non-adiabatic dissipation does not alter the extreme bounds on the coefficient of performance of the power law dissipative Carnot-like refrigerator optimized by both these target functions. The future work will focus on the comparison of different types figure of merit predictions with observed coefficient of performance of real refrigerators working in different dissipation regime.

\end{document}